%% file: article.tex
\documentclass[aps,prb,10pt,floatfix,superscriptaddress,twocolumn,letterpaper,nobalancelastpage]{revtex4-2} 

\usepackage{mycommands}

\begin{document}

    \newcommand{\articletitle}{Weak localization and universal conductance fluctuations\\ in large-area twisted bilayer graphene}
    \title{\articletitle}

    \author{Spenser Talkington}
    \email{spenser@upenn.edu}
    \affiliation{Department of Physics and Astronomy, University of Pennsylvania, Philadelphia, Pennsylvania 19104, USA}

    \author{Debarghya Mallick}
    \affiliation{Materials Science and Technology Division, Oak Ridge National Laboratory, Oak Ridge, Tennessee 37831, USA}

    \author{An-Hsi Chen}
    \affiliation{Materials Science and Technology Division, Oak Ridge National Laboratory, Oak Ridge, Tennessee 37831, USA}

    \author{Benjamin F. Mead}
    \affiliation{Department of Physics and Astronomy, University of Pennsylvania, Philadelphia, Pennsylvania 19104, USA}

    \author{Seong-Jun Yang}
    \affiliation{Department of Chemical Engineering, Pohang University of Science and Technology, Pohang 37673, Republic of Korea}

    \author{Cheol-Joo Kim}
    \affiliation{Department of Chemical Engineering, Pohang University of Science and Technology, Pohang 37673, Republic of Korea}

    \author{Shaffique Adam}
    \affiliation{Department of Physics, Washington University in St. Louis, St. Louis, Missouri 63130, USA}
    \affiliation{Department of Physics and Astronomy, University of Pennsylvania, Philadelphia, Pennsylvania 19104, USA}
    \affiliation{Department of Materials Science and Engineering, National University of Singapore, 9 Engineering Drive 1, Singapore 117575, Singapore}

    \author{Liang Wu}
    \affiliation{Department of Physics and Astronomy, University of Pennsylvania, Philadelphia, Pennsylvania 19104, USA}

    \author{Matthew Brahlek}
    \email{brahlekm@ornl.gov}
    \affiliation{Materials Science and Technology Division, Oak Ridge National Laboratory, Oak Ridge, Tennessee 37831, USA}

    \author{Eugene J. Mele}
    \affiliation{Department of Physics and Astronomy, University of Pennsylvania, Philadelphia, Pennsylvania 19104, USA}

    \date{\today}
    
    \begin{abstract}
        We study diffusive magnetotransport in highly $p$-doped large area twisted bilayer graphene in $1^\circ$, $7^\circ$, $9^\circ$ and $20^\circ$ samples. We report weak localization in twisted bilayer graphene for the first time. All samples exhibit weak localization, from which we extract the phase coherence length and intervalley scattering lengths, and from that determine that dephasing is caused by electron-electron scattering and intervalley scattering is caused by point defects. We observe signatures of universal conductance fluctuations in the $9^\circ$ sample, which has high mobility and is near the van Hove singularity. Further improvements in sample quality and applications to large area moir\'e materials will open new avenues to observe quantum interference effects.
    \end{abstract}
    
    \maketitle
    
    \section{Introduction}\label{section:intro}
        Quantum interference can lead to electronic localization where weak disorder localizes electrons in one dimension while in three dimensions electrons remain delocalized \cite{anderson1958absence,evers2008anderson}. In the scaling theory of electronic localization \cite{abrahams1979scaling}, two dimensions is marginal and the symmetry of the Hamiltonian dictates whether large samples are conductive or insulating \cite{beenakker1997random}. Graphene presents an ideal platform to explore this physics where valley $SU(2)$ symmetry preserving systems are in the symplectic ensemble and are conductive and exhibit weak anti-localization (WAL) in magnetotransport; in contrast, valley symmetry breaking systems are in the orthogonal ensemble and flow to an insulating renormalization group fixed point at long length scales and exhibit weak localization (WL) \cite{das2011electronic}. Previous studies (discussed below) on monolayer graphene and Bernal bilayer graphene have observed WL, WAL, and universal conductance fluctuations (UCF) tuned by defects. Studying quantum transport of twisted bilayer graphene (TBG) in the metallic regime reveals the extent of intervalley coherence \cite{wei2025weak}, and the mechanisms by which phase coherence is lost.
        
        Here we use twist angle as a knob to tune the electronic structure between symmetry classes where at large twist angles there is a conserved valley symmetry and at small twist angles Bragg scattering and hybridization break the valley symmetry. Despite the different electronic structures we observe WL in all samples which we attribute to defects that introduce intervalley scattering. Nevertheless, near the twist angle where the van Hove singularity (vHs) crosses the chemical potential and a crossover between these classes occurs we observe enhanced fluctuations consistent with a crossover between ensembles \cite{adam2002enhanced}, and we analyze the possibility of UCF. Below the vHs, the electronic structure of TBG is well described by two Dirac cones and is in the symplectic class; above the vHs, back-folding and band mixing break the valley symmetry and the system is in the orthogonal class \cite{lopes2012continuum}. To conduct these experiments we leveraged advances in fabrication to create millimeter-sized TBG with $p$-doping far from charge neutrality \cite{yang2019all,yang2022wafer,yang2022twisted}, which ensured that the mesoscopic regime was accessible, all samples were metallic, and that weak localization corrections would be significant. We then conducted temperature dependent magnetotransport measurements.

    \section{Twisted Bilayer Graphene}

        Twisted bilayer graphene is a tunable platform for electronic physics, but to date weak localization has not been observed in TBG, perhaps due to a suppression of WL/WAL near charge neutrality \cite{morozov2006strong}: here we observe weak localization in TBG samples far from charge neutrality. Near the magic angle TBG exhibits electronic localization due to flat bands as first predicted by tight-binding theory \cite{morell2010flat,trambly2010localization} and subsequently explained by low energy theories \cite{bistritzer2011moire,carr2019exact,yudhistira2019gauge,sharma2021carrier,nguyen2021electronic}. These flat bands are now well accepted with direct nano-ARPES measurements \cite{lisi2021observation,utama2021visualization} and observations of localization near magic angle \cite{zhang2020abnormal,gadelha2021localization}. These flat bands enhance the role of electronic interactions leading to correlated insulating states \cite{cao2018correlated} and superconductivity \cite{cao2018unconventional,yankowitz2019tuning} and a plethora of correlated phenomena reviewed in Ref. \cite{andrei2020graphene}. Flat bands are not unique to TBG and arise in a number of distinct physical platforms \cite{morales2016simple,rhim2021singular,talkington2022dissipation,regnault2022catalogue,cualuguaru2022general} leading to similar localization and correlation phenomena. TBG is also notable for its inflated unit cell which compresses the Brillouin zone making Wannier-Stark localization in an electric field possible \cite{wannier1960wave,kolovsky2013wannier,kelardeh2014wannier,iafrate2020bloch,de2024floquet}. Meanwhile at larger twist angles interlayer coherence effects similar to that of Bernal bilayer graphene can become important \cite{shallcross2008quantum,mele2010commensuration,mele2012interlayer} where the band structure becomes highly tunable with electric field \cite{mccann2006asymmetry,castro2007biased,mak2009observation,talkington2023electric} and the electronic structure can be revealed by terahertz optics \cite{talkington2023terahertz,talkington2024linear,mead2025terahertz}. Magnetotransport measurements should provide a detailed probe of this coherence, tunability, and the effect of disorder as explored in a number of recent theory works \cite{wei2025weak,li2024time,guerrero2025disorder,mondal2025magneto}: here we study magnetotransport as a function of twist angle far from the Landau fans measured near charge neutrality.

        To fabricate the samples, monolayer graphene was grown on germanium. The TBG samples were prepared from the monolayers with a previously developed wet transfer technique \cite{yang2022wafer} with the bilayers placed on a sapphire substrate; the total sample dimensions were roughly $5\times 5$ mm. The twist angle uncertainty is estimated as $\pm 1.5^\circ$, consistent with other samples prepared using this technique quantified using transmission electron microscopy diffraction patterns \cite{yang2022wafer}. The Hall effect measurements were performed in a van der Pauw geometry with contacts made using pressed indium wires. The contacts were placed to minimize any effects of tears which could be seen using an optical microscope. Leads were spaced up to 1 mm apart. The measurements were performed in a Quantum Design physical property measurement system from 2 K up to 110 K and over field ranges chosen so that the full set of weak-localization features would be visible.
        
\begin{figure}
            \centering
            \includegraphics[width=\linewidth]{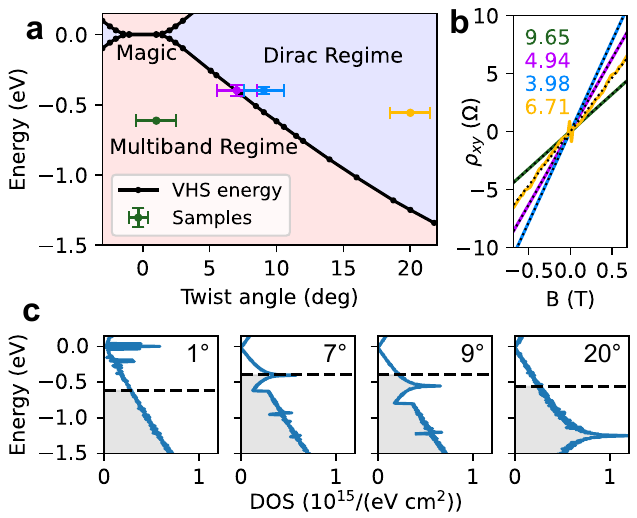}
            \caption{Crossing the van Hove singularity in doped twisted bilayer graphene: the $1^\circ$ sample (green) is deep in the multiband regime, the $7^\circ$ (purple) and $9^\circ$ (blue) samples are near the van Hove singularity, and the $20^\circ$ (orange) sample is deep in the Dirac regime. The twist angle uncertainty is estimated to be $1.5^\circ$. \textbf{(a)} Sample twist and chemical potential plotted with regimes in red and blue with the phase boundary determined from the first van Hove singularity of a tight-binding model. \textbf{(b)} Hall effect measurements at $20~\mathrm{K}$ (colored), linear fits (dotted), and extracted hole densities in $10^{13} /\mathrm{cm}^{2}$. \textbf{(c)} Density of states and chemical potential from fitting hole densities to the tight-binding model density of states.}
            \label{fig:vhs-mu}
        \end{figure}

        As the twist angle decreases, the energy of the vHs decreases as predicted by electronic structure theory \cite{lopes2007graphene,shallcross2013emergent} and observed in tunneling measurements \cite{li2010observation,yan2012angle,wong2015local,kim2016charge} and Raman spectroscopy \cite{ribeiro2015origin}. The vHs separates symplectic ensemble Dirac physics from orthogonal ensemble multiband physics where varying twist angle at fixed chemical potential is expected to tune transport from WAL to WL: this idea underlies the design of the study, as illustrated in Fig. \ref{fig:vhs-mu}(a), where we have manufactured TBG in both regimes and near the vHs with $\sim 0.5$ eV $p$-doping. In the transfer process, some germanium or gold residue may have been introduced and this could explain the $p$-doping. Van Hove singularities where the density of states is drastically enhanced have been long recognized \cite{van1953occurrence} to enable unusual electronic physics including high temperature superconductivity \cite{markiewicz1991van} and correlated states \cite{yuan2019magic,classen2024high}. Monolayer graphene has a vHs at $\sim 3$ eV from charge neutrality \cite{geim2007rise,castro2009electronic} and while doping to this point is predicted to lead to superconductivity \cite{mcchesney2010extended}, drastic measures are needed to reach this level of doping: Rosenzweig et al doped to the van Hove singularity at $5.5\times 10^{14}/\mathrm{cm}^2$ using Yb intercalation and K adsorption \cite{rosenzweig2020overdoping}, but modified the electronic structure in the process. 
        The angle tunable vHs in TBG enables us to cross the vHs and observe enhanced fluctuations \cite{adam2002enhanced} by varying twist angle between samples.

    \section{Weak Localization}\label{section:weak-localization}
        \begin{figure*}
            \centering
            \includegraphics[width=\linewidth]{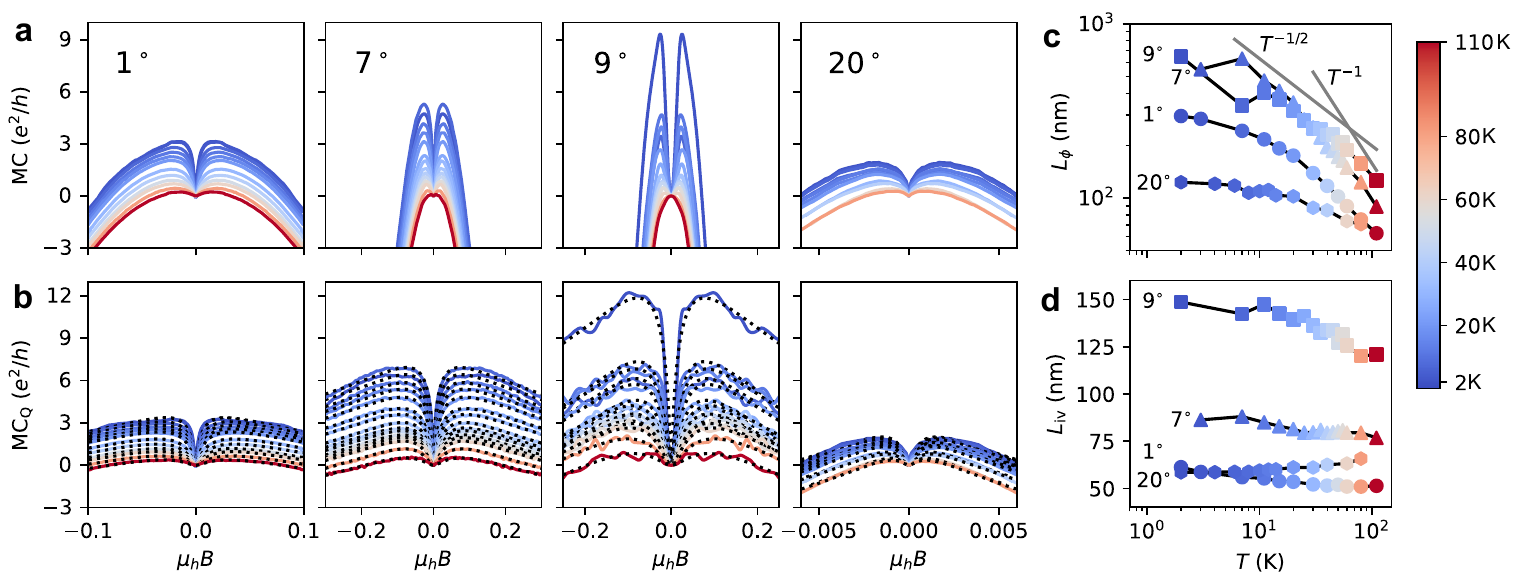}
            \caption{Weak localization in twisted bilayer graphene. \textbf{(a)} Twist-angle and temperature dependent magnetoconductivity (MC) plotted against dimensionless magnetic field $\mu_h B$. In dimensionful units, the $x$-axis limits are from left to right, $\pm$1.20, 0.62, 0.17 and 1.32 T. Note the pronounced weak-localization peak around zero field. The $20^\circ$ sample has low mobility and low MC which is multiplied by 10 to plot on the same axes as the other samples. \textbf{(b)} Quantum contribution to the MC obtained by subtracting the classical Drude term from the total MC; data are fitted (dotted black lines) by the Hikami-Larkin-Nagaoka (HLN) formula for quantum contribution to the MC. \textbf{(c)} Phase coherence length from the HLN fit; power law scaling between $-0.5$ and $-1$ indicates electron-electron scattering is the main mechanism for phase decoherence. \textbf{(d)} Intervalley scattering length from the HLN fit; the roughly temperature independent length suggests that intervalley scattering is caused by point defects in the lattice.}
            \label{fig:weak-localization}
        \end{figure*}
        Before the onset of strong/Anderson localization \cite{anderson1958absence,evers2008anderson}, weak localization occurs as described in the scaling theory of localization \cite{abrahams1979scaling}. A key distinction is between weak localization where constructive interference between time-reversed paths leads to increased localization and weak anti-localization where destructive interference leads to delocalization \cite{kramer1993localization,rammer2004quantum}. In graphene this interference could originate from a 0 Berry phase in the valley symmetry breaking regime (orthogonal) and a $\pi$ Berry phase in the valley symmetry conserving regime (symplectic) \cite{das2011electronic}. Now, applying a magnetic field introduces an additional Aharonov-Bohm phase that modifies the total phase between time-reversed paths leading to decreased interference and hence negative magnetoresistivity (positive magnetoconductivity (MC)) for WL and positive magnetoresistivity (negative MC) for WAL.

        Monolayer graphene with its Dirac cones leading to a quantized $\pi$ Berry phase in the absence of intervalley scattering is a perfect platform to study WAL and WL. Shortly after the isolation of monolayer samples \cite{novoselov2005two}, many experiments observed WAL and WL tuned by the chemical potential and defect structure of samples \cite{morozov2006strong,wu2007weak,tikhonenko2008weak,ki2008inelastic,tikhonenko2009transition,horsell2009mesoscopic,chen2010magnetoresistance,berezovsky2010imaging,moser2010magnetotransport,lara2011disordered,jauregui2011electronic,baker2012weak,hilke2014weak,terasawa2017relationship,terasawa2017temperature,terasawa2019universal}. These experiments were readily understood in terms of WL theory \cite{mccann2006weak,morpurgo2006intervalley,aleiner2006effect,altland2006low,fal2007weak,yan2008weak} and more general theory on disordered transport and electronic properties in monolayer graphene \cite{adam2007self,adam2009crossover,mucciolo2010disorder,adam2012graphene,ping2014disorder}. Extensions to the Bernal bilayer with massive Dirac fermions and Berry phase $2\pi$, where weak localization occurs in the absence of scattering, came quickly with theory works on weak localization \cite{kechedzhi2007weak,kechedzhi2007influence,mccann2013weak} and disordered transport \cite{adam2010temperature} with experimental observations of weak localization coming soon after \cite{gorbachev2007weak,gorbachev2008weak,liao2010gate,yungfu2011negative,liao2011observation,liu2011enhanced,chuang2014weak,icking2025weak}. Despite these successes in monolayer and Bernal bilayer graphene, WL/WAL in TBG has not been observed. We hypothesize that this is because WL is a mesoscopic phenomenon occurring in metallic systems at mesoscopic length scales where diffusive transport takes place. Most previous TBG devices were only available at microscopic scales where ballistic transport predominates, and devices with diffusive transport were near charge neutrality and insulating states where weak localization effects are a small correction \cite{polshyn2019large,sharma2021carrier}. Advances in transfer techniques and ultra-flat substrates have enabled the fabrication of mesoscopic and even macroscopic TBG \cite{yang2019all,yang2022wafer,yang2022twisted}. We leverage these developments to fabricate TBG samples with size up to $5\times 5$ mm in the $p$-doped metallic regime, which enable us to systematically study the diffusive magnetotransport of metallic TBG.

        We take the measured longitudinal ($R_{xx}$) and transverse ($R_{xy}$) magnetoresistance and decompose it into symmetric $\rho_{xx}(B)=(R_{xx}(B)+R_{xx}(-B))/2$ and antisymmetric $\rho_{xy}(B)=(R_{xy}(B)-R_{xy}(-B))/2$ parts to eliminate the effects of sample geometry. Now, we use the Hall effect $\rho_{xy}(B) = B/ne$ to determine the hole density $n$ in the samples, which is of order $10^{13}/\mathrm{cm}^2$ corresponding to highly doped samples well away from charge neutrality. At this chemical potential $K$ and $K_\theta$ are fully mixed, but in the absence of scattering $K$ and $K'$ remain independent pseudospin degrees of freedom. From the conductivity, $\sigma_{xx} = \rho_{xx}/(\rho_{xx}^2+\rho_{xy}^2)$, we can calculate the mobility of holes $\mu_h=\sigma_{xx}(0)/ne$ and estimate the mean free path as $L_\mathrm{mfp} = 2(\sigma_{xx}(0)/(e^2/h))/g_sg_vg_Lk_F$ where $n=\pi k_F^2$ and $g_s=g_v=g_L=2$ are the spin, valley, and layer degeneracies \cite{castro2009electronic,das2011electronic}. We find $L_\mathrm{mfp}$ on the order of 100 nm, which is diffusive for these samples.

        Now the central object of focus in this section is the magnetoconductivity $\mathrm{MC} = \sigma_{xx}(B)-\sigma_{xx}(0)$, which can be fit by a combination of classical and quantum conductivities, $\mathrm{MC}=\mathrm{MC}_C+\mathrm{MC}_Q$, where the classical conductivity is the Drude MC
        \begin{align}
        	\mathrm{MC}_C(B,T) = - \sigma_{xx}(0,T) \frac{(\mu_h(T) B)^2}{1+(\mu_h(T) B)^2}
        \end{align}
        and the quantum conductivity is given by a variant of the Hikami-Larkin-Nagaoka (HLN) model, which includes long-range dephasing terms $B_\phi$ and short-range intervalley scattering terms $B_\mathrm{iv}$ (in other systems $B_\mathrm{iv}$ is caused by spin-orbit coupling) \cite{hikami1980spin,lee1985disordered,chakravarty1986weak,mccann2006weak,das2011electronic,tikhonenko2008weak}. Explicitly, for spin-triplet Cooperons with the same relaxation time
        \begin{align}\label{eq:hln}
            \mathrm{MC}_\mathrm{Q}(B,T) = s \left[F\left(\frac{B}{B_\phi}\right) - 3F\left(\frac{B}{B_\phi+2B_\mathrm{iv}}\right)\right]
        \end{align}
        where $F(z)=\ln(z)+\psi(1/2+1/z)$ for the digamma function $\psi$ and a sample dependent scale factor $s$. The temperature dependent magnetic field scales $B_\phi$ and $B_\mathrm{iv}$ are related to their corresponding ``magnetic" lengths as $L_\phi=\sqrt{\hbar/4eB_\phi}$ and $L_\mathrm{iv}=\sqrt{\hbar/4eB_\mathrm{iv}}$. As in monolayer graphene, spin-orbit coupling is weak relative to intervalley scattering so we neglect the spin-orbit terms.

        In Fig. \ref{fig:weak-localization}, we plot the total MC as a function of temperature and magnetic field for four samples with twist angles of $1^\circ$, $7^\circ$, $9^\circ$, and $20^\circ$. In all samples we see a pronounced cusp at zero field corresponding to WL. While this is expected for the $1^\circ$ sample and possibly the $7^\circ$ and $9^\circ$ samples, the $20^\circ$ sample is expected to have a conserved valley symmetry leading to WAL; the WL in this sample is the evidence that the valley symmetry is broken, and the Berry phase is zero, likely due to the presence of atomic point defects. We expect that defects are particularly prevalent in the $20^\circ$ sample as it has the lowest conductivity and mobility of all samples; to plot the MC on the same axes as the other samples we multiply its MC and MC$_Q$ by 10. By subtracting the classical Drude MC in Fig. \ref{fig:weak-localization}(b) we obtain the quantum MC, which is well fit at low fields by Eq. (\ref{eq:hln}).
        
        Now, we plot the fit parameters in Fig. \ref{fig:weak-localization}(c-d). The temperature dependence of the phase coherence length $L_\phi$ is in the range of $-1/2$ to $-1$, which indicates that phase decoherence is caused by electron-electron scattering; if phase coherence were limited by electron-phonon scattering the phase coherence would fall off faster with $T$ \cite{chakravarty1986weak}. Next, the roughly temperature independent intervalley scattering length indicates that the intervalley scattering mechanism is temperature independent; additionally intervalley scattering from $K$ to $K'$ requires a large momentum transfer corresponding to a scatterer on the scale of the atomic lattice so we conclude that intervalley scattering is caused by atomic point defects.

    \section{Doping Dependence of Weak Localization}

        One may wonder why previous studies on twisted bilayer graphene have not observed weak localization; we suggest that it is because the samples in the diffusive metallic regime were insufficiently doped to observe weak localization. Here we show that the magnitude of weak localization goes as $\sqrt{n}\log(\sqrt{n})$ in Dirac materials: increasing doping enhances weak localization. In the weak magnetic field limit, the amplitude of the weak localization effect for bilayer graphene reduces to the form \cite{beenakker1991quantum}
        \begin{align}
        s = \frac{g_sg_vg_L}{2\pi} \frac{e^2}{h} \frac{m^*}{m_e} \log(1 + \frac{\tau_\phi}{\tau_\mathrm{mfp}}) 
        \end{align}
        where $g_s=g_v=g_L=2$ are the spin, valley, and layer degeneracies, and $m^*=\epsilon_F/v_F^2$ for Dirac materials. Now we note that $\sigma_{xx}(0)=e^2 \nu_2 D$ with the diffusion constant $D=v_F^2\tau_\mathrm{mfp}/2$ therefore
        \begin{align}
        \tau_\mathrm{mfp}
        = \frac{2\hbar}{\epsilon_F} \frac{\sigma_0}{g_sg_v g_L}
        \end{align}
        where $\sigma_0=\sigma_{xx}(0)/(e^2/h)$ and we used the density of states for bilayer graphene, $\nu_2 = (g_sg_vg_L/2\pi)\cdot m^*/\hbar^2$. Next, when electron-electron scattering limits phase coherence (as we have determined is the case here), $\tau_\phi$ is given by \cite{altshuler1982effects}
        \begin{align}
        \tau_\phi = \frac{2\hbar}{k_B T} \frac{\sigma_0}{\log(\sigma_0)}
        \end{align}
        This means that in the low temperature limit (which is even satisfied at room temperature for our samples)
        \begin{align}
        s = \bigg[\frac{g_sg_vg_L}{2\pi}  \frac{\epsilon_F/v_F^2}{m_e} \log\bigg(\frac{g_s g_v g_L \epsilon_F}{k_B T} \frac{1}{\log(\sigma_0)}\bigg)\bigg]\frac{e^2}{h}
        \end{align}
        This expression implies that the weak-localization amplitude becomes large at high density---it scales as $\epsilon_F\log(\epsilon_F)$, or equivalently as $\sqrt{n}\log(\sqrt{n})$---which is precisely the regime of our samples.

        In previous studies of twisted bilayer graphene in the diffusive regime weak localization was not observed \cite{polshyn2019large,cao2020strange}. In Ref. \cite{polshyn2019large}, for a sample at 1.59$^\circ$ the parameters are: $v_F=2.0\times 10^5$ m/s, $\sigma_{xx}(0)=76~e^2/h$, and $n=-2.94 \times 10^{11}/\mathrm{cm}^2$, so $\epsilon_F=4.03$ meV. Meanwhile for our $9^\circ$ sample at 2 K we have $v_F=9.53\times 10^5$ m/s, $\sigma_{xx}(0)=2390~e^2/h$, and $\epsilon_F=-376$ meV. Evaluating the expressions for the 1.59$^\circ$ sample and for our $9^\circ$ sample at $T=2$ K we find $s$ is $0.085~e^2/h$ and $0.715~e^2/h$ respectively. Now, our sample was doped to $n=-3.98\times 10^{13}/\mathrm{cm}^2$. If it had only been doped to $2.9\times 10^{11}/\mathrm{cm}^2$, then $s$ would have been $0.042~e^2/h$ and weak localization would not have been a significant effect. While our $9^\circ$ sample is near the edge of the Dirac regime and close to the van Hove singularity, and the $1.59^\circ$ sample is near the magic regime, we nevertheless expect that this Dirac-only scaling relation captures the key feature of enhanced weak localization at increased density.

    \section{Universal Conductance Fluctuations}\label{section:ucf}
        \begin{figure}
            \centering
            \includegraphics[width=1.04\linewidth]{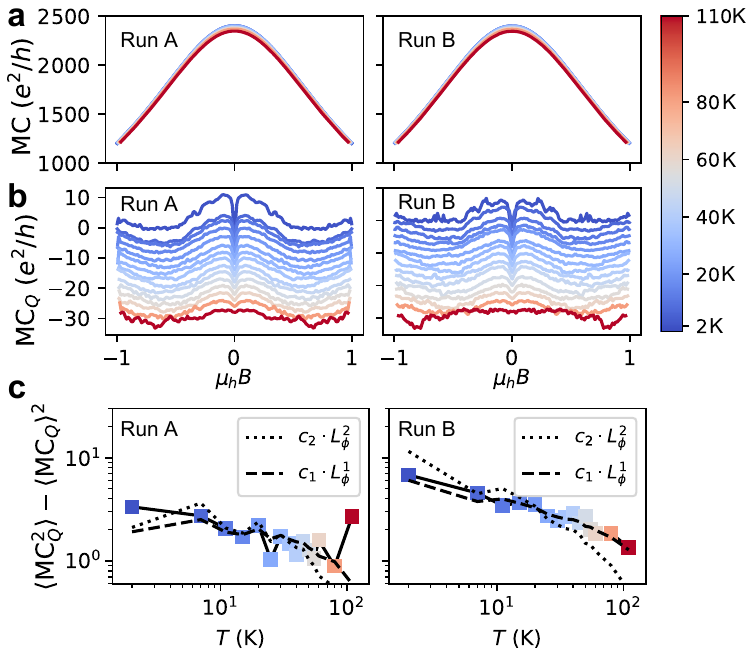}
            \caption{Universal conductance fluctuations in the $9^\circ$ twisted bilayer graphene sample measured with $B$ in the range $\pm 0.7$ T. \textbf{(a)} Total magnetoconductivity (MC) for two separate runs on the same sample. \textbf{(b)} Quantum MC obtained by subtracting classical Drude MC; temperature sweeps are shifted by a constant offset of $2e^2/h$ from each other at $B=0$ for visual clarity. \textbf{(c)} The magnitude of fluctuations is on the scale of the universal conductance $(e^2/h)^2$ and falls off with the phase coherence length $L_\phi$.}
            \label{fig:ucf}
        \end{figure}
        A related phenomenon to weak localization is universal conductance fluctuations (UCF) that often occur in conjunction with each other in diffusive magnetotransport experiments. UCF occurs in systems where coherent scattering of different electronic trajectories happens on the phase coherence length $L_\phi$ \cite{lee1985universal,lee1987universal}. In monolayer graphene, UCF was predicted shortly after predictions for WL \cite{rycerz2007anomalously,kechedzhi2008quantum,kharitonov2008universal,horsell2009mesoscopic,borunda2011imaging,rossi2012universal} and observed in short order, often in the same experiments that demonstrated weak localization \cite{morozov2006strong,staley2008suppression,chen2010magnetoresistance,berezovsky2010imaging,pal2012direct,gopinadhan2012universal,kochat2013universal,terasawa2017relationship,terasawa2017temperature,amin2018exotic,terasawa2019universal,hsieh2021spontaneous}. Likewise, UCF was observed in experiments on Bernal bilayers \cite{gorbachev2007weak,gorbachev2008weak,liao2010current,yungfu2011negative,liao2011observation,chuang2014weak,icking2025weak} and topological insulators with Dirac cones \cite{mallick2021existence}.
        
        To diagnose UCF, we consider fluctuations quantified by the connected correlation function $\langle\mathrm{MC}_Q^2\rangle-\langle\mathrm{MC}_Q\rangle^2$, where the average is taken over magnetic field strengths $B$ and we have subtracted the classical Drude MC. For the right combination of disorder strengths as realized by the $9^\circ$ sample, fluctuations of order $e^2/h$ enter the low temperature MC. Explicitly, UCF are expected to scale as \cite{lee1987universal} 
        \begin{align}\label{eq:ucf}
            \frac{\langle\mathrm{MC}_Q^2\rangle-\langle\mathrm{MC}_Q\rangle^2}{(e^2/h)^2} = c L_*^{4-d}
        \end{align}
        where $c$ is a temperature independent scale factor, $d$ is the effective dimensionality of the channel between the contacts, and $L_*=(1/L_\phi^2 + 1/L_T^2)^{-1/2}$. $L_\phi$ is determined from the HLN model fit to the MC data and the thermal length is $L_T=\sqrt{\hbar L_\mathrm{mfp} v_F/2k_B T}$ with $v_F\approx 10^6$ m/s for the $9^\circ$ sample. With $L_\phi<L_T$, as is the case here, phase decoherence dominates thermal scattering and fluctuations go as $c L_\phi^{4-d}$.

        Now in Fig. \ref{fig:ucf}, we consider two runs of temperature dependent MC measurements. In Fig. \ref{fig:ucf}(a) we see that the gross structure of the MC is the same for the two runs, but once the classical MC is subtracted the quantum MC, Fig. \ref{fig:ucf}(b), the two runs exhibit different structures with fluctuations on the order of $e^2/h$. To quantify the fluctuations, we evaluate $\langle\mathrm{MC}_Q^2\rangle-\langle\mathrm{MC}_Q\rangle^2$ for fields from $\mu_h B=0.2$ to 1.0, where we exclude the low field data since it has a systematic contribution from the weak localization effect. In Fig. \ref{fig:ucf}(c) we see that the fluctuations are on the scale of a few $(e^2/h)^2$ and that the fluctuations fall off with temperature. We fit the data to both $c L_\phi^{2}$ and $c L_\phi^{1}$. The former is expected to be the scaling form in two dimensions, and while it works acceptably at low temperatures it underestimates the fluctuations at high temperatures. This could be due to deviations from square geometry or the presence of point defects not modeled by the theory. Curiously, the fall-off is better captured by $c L_\phi^{1}$ over the full temperature range.
        
        Another unusual feature is that the leads were spaced by $\sim$1 mm which is much longer than $L_\phi\sim 1\ \mu \mathrm{m}$. This is rather surprising as up to logarithmic corrections one expects the amplitude of UCF to go as $L_\phi^2/\mathrm{Area}$ so one would expect that UCF would be attenuated by a factor of $\sim 10^{-6}$ in these samples \cite{lee1987universal}.  The key realization is that this scaling analysis assumes spatial uniformity, but in real systems macroscopic geometry does not necessarily imply many independent coherent regions: a percolation network of very few strong links can dominate electrical transport. In this case, the length to obtain full self-averaging can be much longer than $L_\phi$. This breakdown of the typical scaling has been seen in a number of thin films where UCF was observed at length scales of: 2 mm in In$_2$O$_3$ \cite{milliken1990observation}, 0.1 mm in HgTe \cite{khudaiberdiev2023mesoscopic}, 0.1 mm in Al \cite{delahaye2008coexistence}, 1 mm in Al \cite{delahaye2009electrical}, 0.6 mm in Au \cite{havdala2012ultra}, and 2 mm in Au \cite{belevtsev1996mesoscopic}, all while phase coherence lengths were two to three orders of magnitude smaller. A more exotic mechanism involving many-body coherence and Berry phase effects \cite{varma2026he} has recently been proposed to explain coherent transport at length scales far exceeding the phase coherence length kagome metals \cite{guo2025many}. In our case, the effect of inhomogeneity and a few strong links seems to plausibly explain the persistence of UCF to the mm scale.
        
        Finally, to address why the $9^\circ$ sample exhibits UCF while the fluctuations in the other samples are considerably smaller we point out that the mobility of the $9^\circ$ sample is the greatest of the four samples at nearly 15,000 $\mathrm{cm^2/V\cdot s}$ (see Table. \ref{tab:transport-properties}) and its intervalley scattering length is substantially longer than that of the other samples---Fig. \ref{fig:weak-localization}(d). These features should both help to enable mesoscopic coherence phenomena such as UCF. 

    \section{Discussion}\label{section:conclusion}
        In this study, we report the first observation of weak localization and universal conductance fluctuations in twisted bilayer graphene. While these mesoscopic quantum coherence phenomena have been seen in monolayer and Bernal bilayer graphene for some time, until recently samples of twisted bilayer graphene have either (1) been too small to observe weak localization, (2) been insulating or near to an insulating state where weak localization does not appear, or (3) insufficiently doped for weak localization to be significant. We overcame these difficulties by using a recently developed transfer technique \cite{yang2022wafer,yang2022twisted} to fabricate macroscopic (millimeter-scale) samples with substantial $p$-doping, and we measured their magnetoconductivity. The $p$-doping brought the samples far from charge neutrality, and away from the Landau fan diagrams measured in other experiments of Landau levels and correlated insulators, and into a regime where mesoscopic coherence dominates. This method of sample preparation is much more scalable than typical exfoliation followed by dry transfer and may enable future device applications. Large samples are also desirable for infrared and terahertz optical studies, which could probe low-lying electronic excitations; we have already interfaced terahertz optics with large area twisted bilayer graphene and were able to resolve inter-Landau level transitions \cite{mead2025terahertz}. Looking ahead, the pursuit of even higher quality samples with fewer point defects may enable the observation of mesoscopic physics tuned purely by electronic structure, where the role of defects is secondary, such as weak anti-localization in large angle twisted bilayer graphene with a preserved valley symmetry.

    \section{Acknowledgments}
        We thank Matthew Yankowitz for comments on the manuscript. S.T. acknowledges support from the NSF under Grant No. DGE-1845298. A.H.C., D.M., and M.B. were supported by the U. S. Department of Energy (DOE), Office of Science, Basic Energy Sciences (BES), Materials Sciences and Engineering Division, and the National Quantum Information Science Research Centers, Quantum Science Center. B.M. was partially sponsored by the Army Research Office under Grant Number W911NF-20-2-0166 and W911NF-25-2-0016. S.J.Y. and C.J.K. acknowledge support from the Institute for Basic Science (IBS), Korea, under Project Code IBS-R034-D1. S.A. thanks the University of Pennsylvania for hospitality during his sabbatical, when this work was performed. L.W. acknowledges support from the Alfred P. Sloan Foundation under the award FG-2025-25036. E.J.M. was funded by the Department of Energy under grant DE-FG02-84ER45118.


    \input{references.bbl}

\begin{figure*}
	\includegraphics[width=\linewidth]{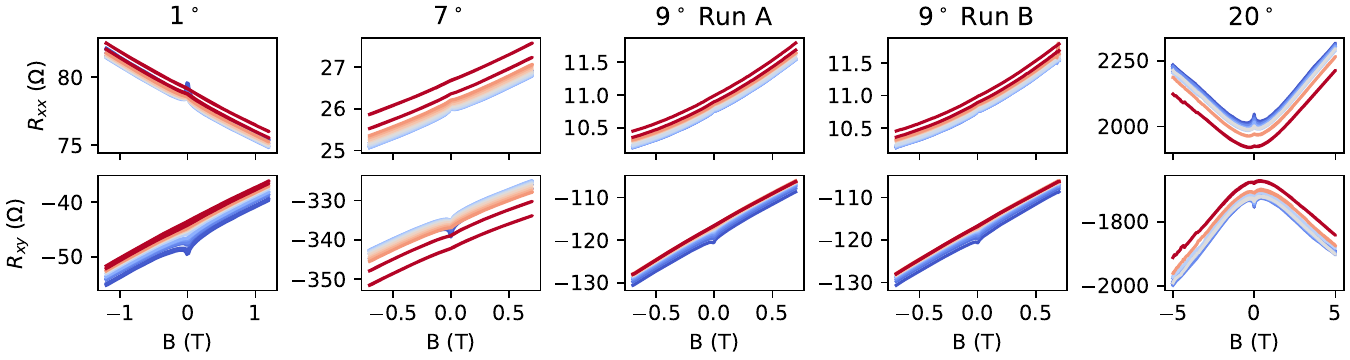}
	\caption{Raw magnetoresistance data. The color bar scale is the same as in the main text, where dark blue is 2 K and dark red is 110 K. Note that the difference between $9^\circ$ Run A and Run B is less than 0.5\% and is not visible until the classical magnetoresistance is subtracted as we do in the main text.}\label{fig:raw-mr}
\end{figure*}

    \newpage

    \appendix

        \section{Raw Magnetoresistance Data}

            Here we plot the raw magnetoresistance data for the $1^\circ$, $7^\circ$, $9^\circ$ run A, $9^\circ$ run B, and $20^\circ$ samples in Fig. \ref{fig:raw-mr}.

        \section{Sample Properties}

            Here we tabulate the sample resistivity, conductivity, mobility, and mean-free paths of the samples.

            \begin{table}[]
                \centering
                \begin{tabular}{c|c|c|c|c}
                    Sample & $\rho_{xx}(0)$ ($\Omega$) & $\sigma_{xx}(0)$ ($e^2/h$) & $\mu_h$ ($\mathrm{cm^2/V\cdot s})$ & $L_\mathrm{mfp}$ (nm) \\\hline
                    $1^\circ$ & 78.4 & 329 & 823 & 74.1\\
                    $7^\circ$ & 26.0 & 992 & 4864 & 313\\
                    $9^\circ$ & 10.8 & 2396 & 14570 & 842\\
                    $20^\circ$ & 2018 & 12.8 & 45.3 & 3.43
                \end{tabular}
                \caption{Sample dependent transport properties at zero magnetic field.}
                \label{tab:transport-properties}
            \end{table}

        \section{Tight-Binding Model}

            For the theoretical model of the density of states in TBG, we used a Slater-Koster tight-binding model of $p$ and $s$ orbitals on carbon atoms with hopping $t(\bm{r}) = V_{pp\pi}(\bm{r}) + V_{pp\sigma}(\bm{r})$ developed and used in Refs. [\onlinecite{trambly2010localization}] and [\onlinecite{moon2013optical}], where the hopping terms are given by:
            \begin{align}
                V_{pp\pi}(\bm{r}) &= t_{pp\pi} e^{-(|\bm{r}|-a_0)/\delta} \left(1-\left(\frac{\bm{r}\cdot \bm{e}_z}{|\bm{r}|}\right)^2\right)\\
                V_{pp\sigma}(\bm{r}) &= t_{pp\sigma} e^{-(|\bm{r}|-d)/\delta}\left(\frac{\bm{r}\cdot \bm{e}_z}{|\bm{r}|}\right)^2
            \end{align}
            for hopping amplitudes $t_{pp\pi} = -2.7$ eV and $t_{pp\sigma} = 0.48$ eV, the unit vector perpendicular to the bilayer is $\bm{e}_z = (0,0,1)$, intra-layer atomic spacing is $a_0 = a/\sqrt{3} = 0.142$ nm, inter-layer spacing is $d = 0.335$ nm, and decay length is $\delta = 0.184 a = 0.0453$ nm. The Dirac point then occurs at $K$ at an energy $E_0=0.7833$ eV, which is an overall shift we add to the Hamiltonian. The atomic positions are obtained through a rigid rotation of two honeycomb lattices by an angle $\theta(m,n)=\mathrm{Arg}[(m\omega^*+n\omega)/(n\omega^*+m\omega)]$; the pair $(m,n)$ parametrizes a commensurate rotation and $\omega=e^{i\pi/6}$. The rotation center is given by a point where two $A$ sublattice sites overlap (no displacement between layers), and this point can be chosen to be the corner of the unit cell.

            For the determination of the van Hove singularities in Fig. \ref{fig:vhs-mu}(a) we calculated the density of states on a $k$-space mesh chosen so that $10^6$ to $10^7$ total states are calculated, for a variety of twist angles from $1^\circ$ to $22^\circ$:
            \begin{center}
                \begin{tabular}{c|c|c}
                    Twist & $(m,n)$ & $E_\mathrm{vHs}$ (eV)\\\hline
                    $1^\circ$   & (33,32) & $0.0021$\\
                    $1.5^\circ$ & (23,22) & $-0.0059$\\
                    $2^\circ$   & (17,16) & $-0.0299$\\
                    $2.5^\circ$ & (14,13) & $-0.0579$\\
                    $3^\circ$   & (23,21) & $-0.0979$\\
                    $5^\circ$   & (7,6) & $-0.258$\\
                    $5.5^\circ$ & (13,11) & $-0.290$\\
                    $7^\circ$   & (26,21) & $-0.402$\\
                    $7.5^\circ$ & (5,4) & $-0.426$\\
                    $8.5^\circ$ & (22,17) & $-0.510$\\
                    $9^\circ$   & (29,22) & $-0.554$\\
                    $10^\circ$  & (19,14) & $-0.618$\\
                    $10.5^\circ$  & (29,21) & $-0.658$\\
                    $11.5^\circ$  & (27,19) & $-0.722$\\
                    $12^\circ$  & (13,9) & $-0.758$\\
                    $14^\circ$  & (20,13) & $-0.890$\\
                    $16^\circ$  & (23,14) & $-1.018$\\
                    $18.5^\circ$  & (9,5) & $-1.182$\\
                    $20^\circ$  & (47,25) & $-1.250$\\
                    $22^\circ$  & (2,1) & $-1.342$
                \end{tabular}
            \end{center}

            For the densities of states plotted in Fig. \ref{fig:vhs-mu}(c) we used $(m,n)=$ $(33,32)$, $(26,21)$, $(29,22)$, $(47,25)$ for the $1^\circ$, $7^\circ$, $9^\circ$, and $20^\circ$ samples, respectively.
            To find the chemical potential, we integrate the density of states to the energy where the number density equaled the number density found through Hall effect measurements in Fig. \ref{fig:vhs-mu}(b).
            
            The twist angle variability across the sample was estimated as $\pm 1.5^\circ$ which could influence the chemical potential. To quantify this effect, we calculated the chemical potential for the $1^\circ$ sample density at a $2.5^\circ$ twist, $(14,13)$; the $7^\circ$ sample at $5.5^\circ$ / $(13,11)$ and $8.5^\circ$ / $(22,17)$; the $9^\circ$ sample at $7.5^\circ$ / $(5,4)$ and $10.5^\circ$ / $(29,21)$; and the $20^\circ$ sample at $18.5^\circ$ / $(9,5)$ and $21.5^\circ$ / $(2,1)$. The uncertainty in the data points in Fig. \ref{fig:vhs-mu}(a) is the variability among the chemical potentials determined for the same sample.

\end{document}

%% file: references.bbl
%